# From cryptomarkets to the surface web: Scouting eBay for counterfeits

Felix Soldner[1], Fabian Plum[2], Bennett Kleinberg[3,4], Shane D Johnson[3]

[1]GESIS – Leibniz Institute for the Social Science, Köln, Germany
[2]Department of Bioengineering, Imperial College London, UK
[3]Dawes Centre for Future Crime, Department of Security and Crime Science, University College London, UK
[4]Department of Methodology and Statistics, Tilburg University, The Netherlands

Corresponding author: Felix Soldner, felix.soldner@gesis.org

## Abstract

Detecting counterfeits on online marketplaces is challenging, and current methods struggle with the volume of sales on platforms like eBay, while cryptomarkets openly sell counterfeits. Leveraging information from 453 cryptomarket counterfeits, we automated a search for corresponding products on eBay, utilizing image and text similarity metrics. We collected data twice over 4-months to analyze changes with an average of 159 eBay products per cryptomarket item, totaling 134k products. We found identical products, which would warrant further investigation as to whether they are counterfeits. Results indicate increasing difficulty finding similar products over time, moderated by product type and origin. Future improved versions of the current system could be used to examine possible connections between cryptomarket and surface web listings more closely and could hold practical value in supporting the detection of counterfeits on the surface web.

## 1. Introduction

Big online shopping platforms (e.g., eBay, Amazon) and social media platforms (e.g., Instagram, Facebook) struggle to deal with the increase in counterfeit sales on their platforms (BBC, 2015; Conlon, 2017; Ihaza, 2017; Mooij, 2018; Scheck, 2019; Suthivarakom, 2020). Counterfeits are physical or digital goods that violate intellectual property (IP) rights (e.g., copyrights, trademarks) (OECD/EUIPO, 2019; WTO, 1994), and current measures seem insufficient to deter counterfeit sales (Duhigg, 2019; Zimmerman, 2020). With the increase in counterfeit sales, which are detrimental to brand values and can hurt customers financially and physically (EMCDDA-Europol, 2017), the reliable and efficient detection of counterfeits on online shopping platforms has become increasingly important.

Aside from surface web markets (i.e., eBay, Amazon), counterfeits are also sold on cryptomarkets – online shopping platforms on the deep web – which have received increasing attention from the research community and law enforcement (Baravalle & Lee, 2018; Christin, 2013; Europol, 2017; Ghosh et al., 2017; Van Buskirk et al., 2016). Most markets utilize The

Onion Router (Tor) network, which directs internet traffic through a relay network, ensuring a high degree of anonymity for vendors and customers (Çalışkan et al., 2015; Gehl, 2018; The Tor Project, Inc., 2020). The most commonly sold goods are drugs (Rhumorbarbe et al., 2016; Soska & Christin, 2015), but weapons, phishing information, hacking services, counterfeits, and more are also offered (Douglas, 2015; Roberts & Hernandez-Castro, 2017; van Wegberg et al., 2018). Apart from counterfeits, almost none of cryptomarkets products or services would be offered on online shopping platforms on the surface web. While counterfeits are typically sold (deceptively) as genuine products on the surface web, they are sold openly as counterfeits on cryptomarkets. Given that counterfeits are present on both cryptomarkets and surface web markets, it seems plausible that activity on both markets might be interdependent. Although previous reports on counterfeits have noted that sellers seem to operate across cryptomarkets and surface web markets (EMCDDA-Europol, 2017; Europol, 2017), the extent to which this is the case and for which products are unknown. Similarly, individuals purchasing counterfeits on cryptomarkets might resell the items for profit on surface web platforms, or vendors on cryptomarkets might conduct market research on surface web platforms (e.g., forums, shopping platforms) to find out which products are in high demand to determine which products they should counterfeit and offer.

Because counterfeits on cryptomarkets are sold openly, this presents an opportunity to use the available information (product names, descriptions, images, etc.) to search for matching listings on the surface web. This work explores how an automated search for counterfeits on eBay might work based on current cryptomarket counterfeits. By utilizing text and image similarity metrics between cryptomarket and surface web listings, as well as a ranking system of the similarity scores, we determine the best matches, which could subsequently be prioritized for manual inspection. Since authorities can often only react to incidents of fraud or are faced with intensive manual investigative web searchers to find counterfeits (FBI, 2018), the proposed system aims to demonstrate how manual searches for counterfeits could be supported with automated approaches to become more efficient. Specifically, we use cryptomarket counterfeit product names to search for and collect eBay product information twice within four months to enable an examination of how offers change over time. We then determine similarities between cryptomarket and eBay listings by merging automatically generated and human-annotated similarity scores. Product matches are then ranked by similarity and manually inspected to determine if we can find the same products across markets. Thus, this study explores how manual time-intensive investigations for finding counterfeits on the surface web could be supported through partial automation of the search process.

Our system does not use a supervised classification approach since we have no ground truth data (i.e., eBay listings labeled as counterfeits) required for training a model, and we also do not know how many or if any cryptomarket products can be found on eBay. Lastly, supervised classification systems tend to only work well within the domain they are trained in and do not generalize well (Geirhos et al., 2020). Thus, the performance of a classifier trained to detect fake watches might not be transferable to other products, such as shoes, mobiles, or clothes. Given the reasons above, we devised a method that can work across various products, is not reliant on annotated data, and circumvents the problem of not knowing the true distribution of counterfeits on eBay.

## 2. Data

We first collected cryptomarket (CM) listings of counterfeits, which we subsequently searched for on eBay. We manually collected 453 CM listings on 25 June 2021, while eBay data was collected twice within four months. 66,430 eBay listings were collected automatically between 28 June and 1 July 2021 (period 1), and 68,532 were collected between 5 and 9 November 2021 (period 2). During each eBay scrape, we obtained text data and image links, which we used to collect high-resolution images separately (data is available upon request). The Ethics Committee of the Department of Security and Crime Science, University College London, approved the study and data collection.

### 2.1. Cryptomarket data

Using the Tor browser, we visited 12 cryptomarkets listed on "www.dark.fail" and determined if they contained counterfeits. We selected Darkode, Torrez, White House Market, and World Market because they contained the most counterfeits, which were explicitly categorized as counterfeits. The information on cryptomarket counterfeits was manually collected by locally saving the web pages and images. Automated data collection from anonymity networks can be challenging and is known to lead to data gaps (Ball et al., 2019; Du et al., 2018; Van Buskirk et al., 2015). Since we only needed a small amount of cryptomarket data, we favored a manual approach to ensure reliability. The saved HTML pages were parsed using the Python package "selectorlib" (Rajeev, 2019). We collected information from the newest 50 counterfeits (if available) in each product category to increase the likelihood of finding current matches on the surface web. We collected information on 453 counterfeits, covering five product categories (Table 1).

| Category | Cryptomarket | | | | Total | % |
| --- | --- | --- | --- | --- | --- | --- |
| | Darkode | Torrez | W. H. Market | World Market | | |
| Clothes | - | 44 | - | - | **44** | **9.71** |
| Electronics | - | 1 | 28 | - | **29** | **6.40** |
| Jewellery | - | 44 | 50 | 50 | **144** | **31.79** |
| Other | 34 | - | 35 | 29 | **98** | **21.63** |
| Watches | 50 | 39 | 49 | - | **138** | **30.46** |
| **Total** | **84** | **128** | **162** | **79** | **453** | **100.00** |

Table 1. Counterfeits for each market across categories. "Other" contains products such as wallets, (hand) bags, and hats, but items that are also found in existing categories, such as shirts, shoes, or jewellery, as categorization varied by markets.

For each CM product, we collected the vendor's name, vendor details, product title, description, price (USD), origin, and (possible) shipping destination(s). Furthermore, we collected 1,488 images from these counterfeit listings, with the majority containing four images (Mean=3.54, Median=4, SD=1.25, range=1-5). Figure 1 shows an example listing scraped from Darkode.

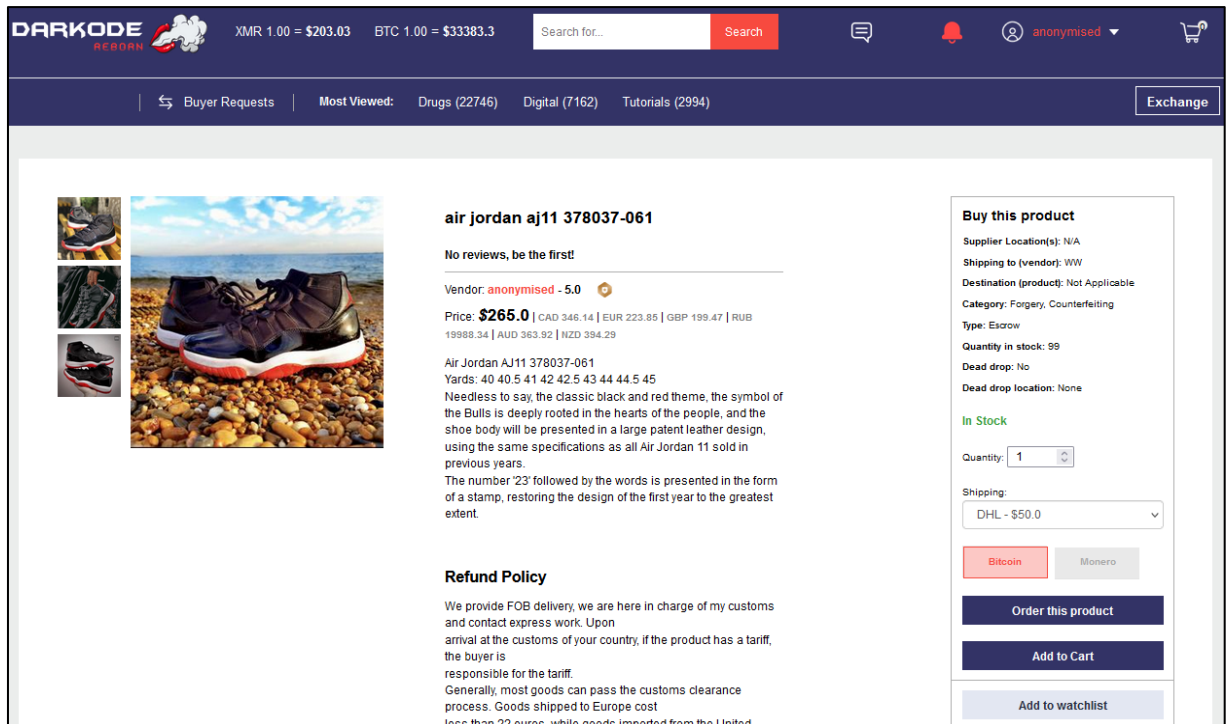

Figure 1. Screenshot of a counterfeit listing on Darkode.

## 2.2. eBay data

To collect eBay data, we automated a product search based on the counterfeit product names and scraped the first page of the results, equating to approximately 200 listings per search. When generating the search terms, we manually removed any words that would indicate that a product was a counterfeit (e.g., "fake", "replica", "counterfeit") as we would not expect such terms to be included in the open web adverts. To scrape eBay product information, we used the Python package "selectorlib" (Rajeev, 2019). We used "www.eBay.com" to search for and collect product information. All scraped information was publicly accessible and obtained without an account. Each of the eBay scrapes (period 1-2) involved three steps:

I. Searching for products on eBay using counterfeit product names and scraping the product links on the first result page.
II. Scraping detailed product information and obtaining the associated image links through the product links.
III. Scraping all images through their links but altering them to obtain images in their native resolution. The number of images was limited to a maximum of 10 per listing.

For the first eBay scrape period, we collected 66,430 listings and found, on average, 156 listings for each search (i.e., for every CM listing product name) (Median=200, SD=79, range=1-221). For the second eBay scrape period, we collected 68,532 listings and found, on average, 162 listings for each search (Median=200, SD=69.85, range=2-252). For period 1, for 27 CM products, no eBay products were found, and for period 2, no eBay products were found for 29 CM products.

For each eBay listing, we collected the product title, specifics (e.g., height, weight), descriptions, price (USD), origin, shipping destination(s), return policy, condition, assurance policies (e.g., money-back guarantee, authenticity check), as well as the vendor's name, feedback score, and the positive percentage of the feedback score (aggregated by eBay). On average, we scraped seven images for each eBay listing (median=7, SD=2.72, range=1-10) and obtained a total of 935,100 unique images across both scrapes.

### 2.3. Cryptomarket and eBay descriptive statistics
#### 2.3.1. Product origins and destinations

Figure 2 shows the product origins for CM and eBay products for the first and second scrape periods. In line with other findings (Europol, 2017), most counterfeits seem to have originated in Hong Kong and China, followed by the UK. In contrast, most eBay listings seem to originate from the USA, which is expected since we used the USA eBay location, followed by Japan. Most product destinations were indicated as "worldwide" or were undeclared (CMs), providing only limited information.

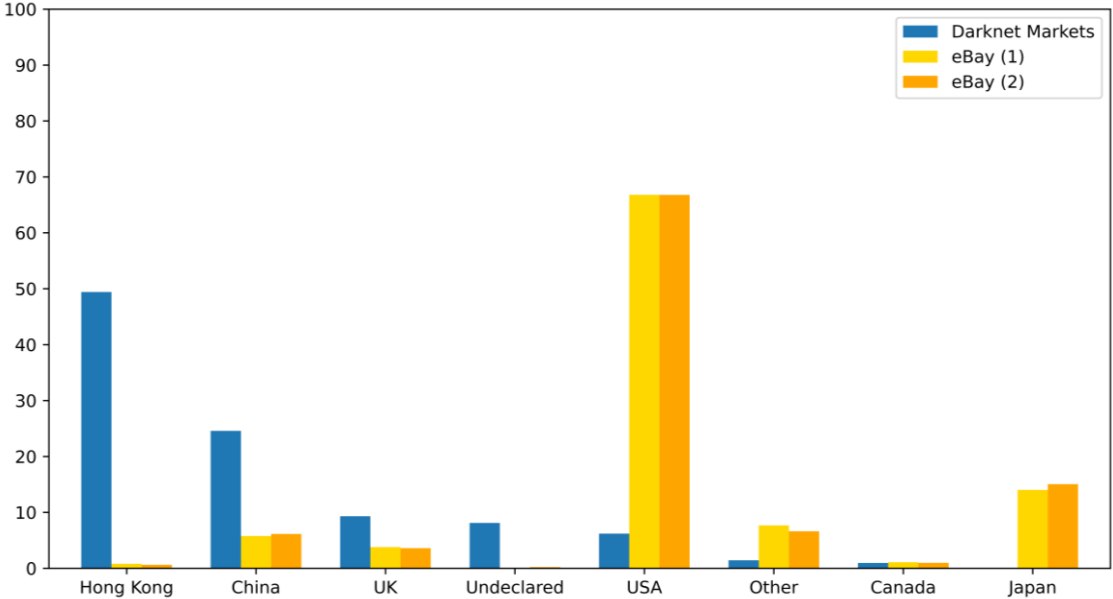

Figure 2. Percentage of product origins in. Origins contributing less than 1% of total products are aggregated into "Other".

#### 2.3.2. Prices

Table 2 shows descriptive statistics for the price of CM and eBay products. CM product prices range from 15 to over 6,000 USD, with electronic products having the highest average prices. Most eBay listings were for immediate purchase. However, customers could occasionally bid on them for which a minimum price was advertised (e.g., 1 cent). Similar to CM products, most categories contained some eBay products with extremely high prices of up to 999k USD for particular watches.

| Category | Price (USD) | | | | | | | | | | Listings | |
|---|---|---|---|---|---|---|---|---|---|---|---|---|
| | min | | max | | median | | mean | | SD | | | |
| | CM | eBay | CM | eBay | CM | eBay | CM | eBay | CM | eBay | CM | eBay |
| Clothes | 67 | 0.01 | 290 | 124,999 | 119 | 35 | 145 | 175 | 63 | 1,488 | 44 | 15,043 |
| Electronics | 140 | 0.01 | 6,250 | 13,000 | 1,275 | 550 | 1,687 | 689 | 1,837 | 752 | 29 | 7,807 |
| Jewellery | 40 | 0.01 | 3,215 | 650,000 | 75 | 350 | 198 | 5,613 | 387 | 1,9627 | 144 | 42,355 |
| Other | 15 | 0.01 | 5,000 | 45,000 | 150 | 196 | 279 | 464 | 625 | 982 | 98 | 33,493 |
| Watches | 230 | 1.63 | 1,480 | 999,999 | 410 | 3,703 | 488 | 17,829 | 242 | 55,456 | 138 | 36,264 |
| Average/Total | 98 | 0.33 | 3,2467 | 366,600 | 406 | 967 | 559 | 4,954 | 631 | 15,661 | 453 | 134,962 |

Table 2. Product price distributions across markets and categories.

## 3. Similarity metrics

We calculated similarity scores between the listings to identify products on eBay that resemble the counterfeits sold on the CM. Specifically, we calculated text and image similarity scores between the CM listings and their associated eBay search results to capture different aspects of similarity.

### 3.1. Text similarities

To compare the texts between CM and eBay listings, we calculated the *Word Mover Distance* (Kusner et al., 2015) and three *cosine similarities* between the titles and descriptions of CM and eBay listings (see Appendix 1 or descriptive statistics on word occurrences in the texts). We also generated metrics, such as the Levenshtein distance (Levenshtein & others, 1966) or the Jaccard index (Jaccard, 1912; Ríssola et al., 2020), based on q-grams and document embeddings but removed them since they showed strong correlations (>0.7) with other similarity measures. Since eBay listings contained three types of product descriptions (item specifics, eBay description, and seller description), we merged them before calculating any similarity score. For all text similarity metrics and any text pre-processing steps, we used the Python package "spaCy" (Honnibal & Montani, 2017).

#### 3.1.1. Word Mover Distance

Calculation of the Word Mover Distance (WMD) requires the use of word embeddings. These represent words in a vector space, in which semantically similar words are closer to each other than semantically dissimilar words (Jurafsky & Martin, 2019). We used the pre-trained Word2Vec embedding space trained on the Google News dataset to create word embeddings for each document and used the Python package "genism" to calculate the WMD (Kusner et al., 2015; Pele & Werman, 2008, 2009). The WMD score indicates the minimum cumulative (Euclidian) distance the word embeddings of document A have to travel to the word embeddings of document B within the embedding space (Kusner et al., 2015). A WMD score of 0 would indicate no distance between the compared documents, indicating the highest similarity. Any WMD score greater than 0 indicates a distance, with greater scores indicating a larger distance between the documents and, hence, less similarity between them. Before calculating the WMD, we removed all stop words from each document and made all text lowercase.

### 3.1.2. Cosine Similarity

**Q-grams:** Q-grams are character-based strings of length q (Ukkonen, 1992). In our case, we decided to split each document into character lengths of 3 (e.g., words such as "good" into "goo" and "ood"). By creating q-gram frequency vectors for each document, we can calculate the cosine distance between the documents.

**S-BERT:** Next to the Euclidian distance in the WMD, we also calculated the cosine similarity between document embeddings, instead of word embeddings, using the neural network language model Sentence-BERT (Reimers & Gurevych, 2019), a modification from BERT (Bidirectional Encoder Representations from Transformers) (Devlin et al., 2019). For our use case, we choose the model instance **"paraphrase-MiniLM-L6-v2"[1],** which is finetuned with 12 datasets[2] because it scores highly on several benchmark datasets[3] and prioritizes fast processing. Before creating the document embeddings, we made the text lowercase but omitted other pre-processing steps, such as stop word removal, as the model is trained on unchanged texts.

**Universal Sentence Encoder:** We calculated the cosine similarity between document embeddings generated with the Universal Sentence Encoder (Cer et al., 2018). We utilized a pre-trained model that used the deep averaging network (DAN) architecture. The model was trained on various texts from Wikipedia, other web resources, and the Stanford Natural Language Inference corpus (Bowman et al., 2015). We implemented version 4 of the model with TensorFlow (Abadi et al., 2016).[4] Similar to generating S-BERT embeddings, we first made all text lowercase but omitted other pre-processing steps.

### 3.2. Image similarities

We combined several comparison methods to compute image similarity metrics, including color histogram correlations with noise removal, different feature extractors and matching algorithms, and a custom-built and trained Siamese deep neural network. We produced similarity scores for every image of each cryptomarket listing (1588 images in total) compared to every image of each surface web listing of the respective search query, resulting in a total of ~3.5 million comparisons with five similarity scores each. Most scores are computed with functions native to the "OpenCV" Python package (Bradski, 2000). We decided to select only the maximum scores obtained for any of the image comparisons for each metric.

### 3.2.1. Image pre-processing

Cryptomarket images were rescaled to a maximum resolution of 1024 x 1024 pixels to preserve aspect ratios, facilitate faster processing, and encourage comparable feature sizes (see Appendix 2 for more details). Since some image comparing methods are susceptible to changes in aspect ratio, non-square images were padded (i.e., adding black borders vertically or horizontally). All files were stored in .jpg format with lossless compression to minimize the influence of downsampling and compression artifacts. The eBay images were not downsampled for low-level comparisons (i.e., color histogram comparison and feature matching). However, for the custom-built Siamese network (implemented in Keras), eBay images were padded and downsampled to 299 x 299 pixels.

---

[1] Model details: https://huggingface.co/sentence-transformers/paraphrase-MiniLM-L6-v2
[2] Model types and performances: https://www.sbert.net/docs/pretrained_models.html
[3] Benchmark comparisons: https://www.sbert.net/_static/html/models_en_sentence_embeddings.html
[4] Model details: https://tfhub.dev/google/universal-sentence-encoder/4

### 3.2.2. Color histograms

We compared color histograms of bilaterally blurred images (Tomasi & Manduchi, 1998) of cryptomarket and eBay images with a kernel size of $5\ x\ 5$ to counter the influence of image noise and compression artifacts on resulting similarity scores. Histograms are produced for each RBG colour channel and normalized before comparison via histogram correlation.

### 3.2.3. Feature detection & matching

We compared the cryptomarket and eBay images by matching feature descriptors generated by three feature detectors: *Scale Invariant Feature Transform* (SIFT) (Lowe, 1999, 2004), *Speeded Up and Robust Features* (SURF) (Bay et al., 2008), and *Oriented fast and Rotated Brief* (ORB) (Rublee et al., 2011). These image *features*, such as corners, blobs, and edges, can be matched between different images. The feature detectors differ in processing speed and ability to handle changes in scale, rotation, distortion, and illumination (Karami et al., 2017; Tareen & Saleem, 2018).

1,000 features were extracted from each image and matched by L2 distance, generating two highest-ranking matches per feature. Likely matches were then filtered, as suggested by Lowe (2004). To rank extracted descriptors by their likelihood, we used the *Fast Library for Approximate Nearest Neighbors* (FLANN) (Muja & Lowe, 2011). We then divided the number of likely matches by the number of extracted features to normalize similarity scores for each image pair. In rare cases of highly repetitive textured backgrounds, the number of repeatedly matched features can exceed the number of extracted features, resulting in scores higher than 1.

### 3.2.4. Siamese neural network

We also trained a Siamese neural network with two identical Inception v3 networks (Szegedy et al., 2015, 2016) as the convolutional pathways with frozen weights and pre-trained on the ImageNet dataset (Deng et al., 2009). The detailed architecture can be found in Appendix 3.

Since the dataset lacks ground truth for likely matches between cryptomarket and eBay listings, but we require such matches for product image pairs to train the Siamese network, we treat all images within the same listing as likely matches and others as unlikely matches. We created "triplets" from two images, one from cryptomarket and one from the eBay listings, with a binary label indicating whether they belonged to the same listing. A positive triplet contains two images of the same product, and a negative contains two images of different products. We randomly selected positive and negative samples to ensure balanced training, resulting in 3,393,154 triplets, with 80% used for training and 20% for validation. The final trained network reached an accuracy of 82.95% and generated a single similarity score ranging from -1 (low similarity, high confidence), over 0 (low confidence), to 1 (high similarity, high confidence). A detailed procedure for calculating the similarity score can be found in Appendix 4.

### 3.3. Determining similarity between product pairs

While generating multiple metrics helps capture different aspects of similarities, incorporating them into a single meaningful score poses an additional challenge. Instead of adding all similarity scores, we determined an individual weight for each metric. To do so, we asked crowd workers to annotate 1000 listing pairs, rating their overall similarity (rating 1 to 7), and ran a regression analysis with the similarity metrics as the independent variables and the human-annotated similarity score as the dependent variable. We inferred weights for each similarity metric, which we used to generate the final similarity scores for all unannotated listing pairs. The annotation task was reviewed by the ethics committee of the UCL Department of Security and Crime Science and was exempted from requiring approval by the central UCL Research Ethics Committee. Participants provided informed consent before taking part in the study.

#### 3.3.1. Sampling annotation data

We generated a preliminary manual scoring and ranking procedure to ensure that sampled product pairs contained a broader range of similarity scores. We scored product pairs by equally weighting image and text similarity and merging them by cumulating unusual score distribution counts, indicated by scores of larger or lower than two standard deviations from the metric mean. Ranking the product by similarity, we then sampled the first and last 250 products and a random sample of 500 from the remaining products, resulting in 1000 sampled product pairs (for a detailed description, see Appendix 5).

#### 3.3.2. Annotating similarity scores

We recruited 220 participants from the crowdsourcing platform Prolific and redirected them to a Qualtrics survey in which each annotated ten listing pairs, enabling us to obtain at least two similarity ratings for each pair. Participants were presented with the HTML page of a cryptomarket and eBay listing for each product pair we hosted locally. Distracting information irrelevant to the product (e.g., advertising, recommended listings) was omitted, and usernames were anonymized. For each viewed pair, we asked participants: "Based on the combination of images and texts, how likely are the listings about the exact same product?". Participants rated the question on a 7-point Likert scale with the labels 1: "Not at all", 4: "Somewhat", and 7: "Completely". Figure 3 shows the distribution of similarity ratings, including the first and second ratings of the same product matches.

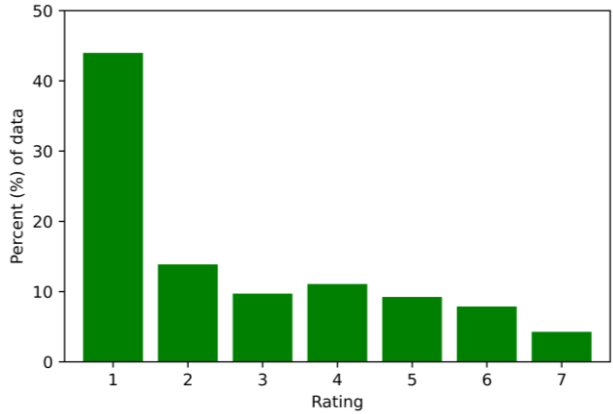

Figure 3. Distribution of annotated similarity ratings.

To examine the extent to which participants agreed on their similarity ratings between cryptomarket and eBay listings, we calculated the associated linear weighted Cohen's Kappa score for the image, text, and overall similarity ratings (Table 3). Annotators seem to agree only slightly (0.01-0.20) or fairly (0.21-0.40) with each other (Cohen, 1960).

| Similarity rating | Product group (N) | | | | | |
|---|---|---|---|---|---|---|
| | Clothes (94) | Watches (136) | Other (259) | Jewellery (394) | Electronics (65) | All (948) |
| Images | 0.07 | 0.22 | 0.27 | 0.15 | 0.21 | 0.20 |
| Texts | 0.06 | 0.12 | 0.15 | 0.10 | 0.36 | 0.14 |
| **Overall** | **0.11** | **0.07** | **0.25** | **0.11** | **0.24** | **0.16** |

Table 3. Similarity rating agreements between listings for each product group and overall.

### 3.3.3. Predicting similarity scores

Although the agreements between annotators were low, we decided to examine the ratings and compare them to the hand-crafted rating system we used to sample the annotation data. We averaged the similarity ratings for each product pair and performed a regression analysis with the averaged human ratings as the dependent variable and the automated generated similarity metrics as independent variables. To assess multicollinearity between independent variables, we calculated Variance Inflation Factors (VIF) and excluded the word mover distance (product names), ensuring that all VIF scores were below 5, meeting acceptable intercorrelation levels (Craney & Surles, 2002; Kim, 2019). We then trained and tested a Random Forest regressor (RFR), an ordinary least squares linear regression model (OLS), and a linear Support Vector regression (SVR) model, each with a 10-fold cross-validation procedure. All MAPE scores are high, with the SVR model performing slightly better with a MAPE (std) of 49.48 (4.02). We used the Python package "scikit-learn" (Pedregosa et al., 2011) for the analyses and the default settings for each model (see Appendix 6 for model settings). We used the product pair ranks obtained with the SVR model for any further analyses. Table 4 shows the SVR metric coefficients, indicating the strength and direction of how the metrics influence the final predicted similarity score. Only USE (product names), q-grams (descriptions), and product types were significant at the 0.05 level.

| Category | Metric | Coef. |
|---|---|---|
| Text (Product names) | Cosine (dist) Q-grams (3) | -0.864 |
| | Cosine S-BERT | -0.148 |
| | Cosine USE | 0.957* |
| Text (Descriptions) | Cosine (dist) Q-grams (3) | -1.802* |
| | Cosine S-BERT | 0.002 |
| | Cosine USE | 0.932 |
| | WMD | -1.540 |
| Images | Histogram (blurred) | -0.106 |
| | ORB | -1.580 |
| | SIFT | 0.115 |
| | SURF | -0.498 |
| | Siamese (merged score) | 0.284 |
| Product Type | Dummy variable | 0.132* |

Table 4. SVR coefficients; Significance level: * = $p < 0.05$.

## 4. Examining similarity scores and product matches

Before examining the product matches, we excluded 643 pairs of the first and 647 pairs of the second scrape period since they lacked image similarity scores due to missing cryptomarket or eBay images. In the coming sections, we present findings as to how the similarity scores varied across product categories, product origins, and eBay scrape periods to understand their importance in affecting similarity scores, which could be valuable in determining which eBay products are more likely to be found identical to cryptomarket counterfeits. We also manually inspected 200 product matches to assess how well the matching and ranking procedures work in finding identical products across the cryptomarkets and eBay.

### 4.1. Changes in listing similarities across categories, origins, and scrape periods

We ran a three-way ANOVA with the similarity metric as the dependent variable to examine whether the similarity scores varied over product categories, product origins (eBay), scrape periods and whether there were any interactions. All main effects of product categories ($F(4)$ = 5489.36, $p < 0.001$), product origins ($F(5) = 1259.25$, $p < 0.001$), and scrape periods ($F(1)$ = 1087.55, $p < 0.001$) were statistically significant. Two-way interactions between product categories and product origins ($F(20) = 203.01$, $p < 0.001$), product categories and scrape periods ($F(4) = 78.98$, $p < 0.001$), as well as product origins and scrape periods ($F(5) = 16.29$, $p < 0.001$), were also statistically significant, as was the three-way interaction between all factors ($F(20) = 7.30$, $p < 0.001$).

Since we were interested in how product categories and product origins differ, as well as how they change across scrapes, we performed post-hoc t-tests between all product categories and product origins for each scrape period as well as between scrape periods. Table 5 and Table 6 show the Cohen's *d* effect sizes between all product categories and product origins. For both tables, the lower-left diagonal half represents the differences within the first scrape period and the upper-right diagonal half within the second. The direction of Cohen's *d* values can be read from row to column. Specifically, each table cell indicates the similarity difference between the reference category (row name) and the target category (column name). For example, the cell (Table 5) with the value of $d = -0.31$ for electronics (row) and watches (column) indicates the similarity difference from electronics to watches within the first scrape period. Looking at the similarity difference between the same categories within the second scrape period, we look at the electronics column and watches row with *d* = 0.37. Since the column and row labels are flipped, the value of *d* = 0.37 indicates the similarity difference from watches (row) to electronics (column). Thus, if the effect size sign (+, -) flipped from the first to the second scrape period (or vice versa) for the same categories, the direction of similarity difference is the same in both periods. We used the Bonferroni alpha level correction to account for multiple comparisons with a starting alpha level of 0.05, resulting in an adjusted alpha level of 0.0016.

|             | Watches | Other  | Electronics | Jewellery | Clothes |
|-------------|---------|--------|-------------|-----------|---------|
| **Watches**     | -       | -0.68* | 0.37*       | -0.14*    | -1.73*  |
| **Other**       | 0.48*   | -      | 0.93*       | 0.45*     | -0.85*  |
| **Electronics** | -0.31*  | -0.62* | -           | -0.42*    | -1.91*  |
| **Jewellery**   | 0.24*   | -0.19* | 0.44*       | -         | -1.22*  |
| **Clothes**     | 1.67*   | 0.76*  | 1.55*       | 0.96*     | -       |

Table 5. Cohen's *d* effect sizes for product category comparisons (row to column) for the first (lower-left diagonal half) and second (upper-right diagonal half) scrape period. Significance level: * = p < 0.05 (Bonferroni corrected).

Similarity changes between product categories were all significant across both scrape periods (Table 5). In the first and second scrape period, we can see the strongest Cohen's *d* differences between Clothes and Watches, and between Clothes and Electronics. Cohen's *d* differences from the first to the second scrape period mostly became stronger, except for the differences between Jewellery and Watches as well as Jewellery and Electronics. The direction of differences did not change across scrape periods. Clothes seemed to show the lowest while Electronics had the highest similarities across scrape periods.

Similarity differences between product origins were almost all significant (Table 6), except for the difference between Europe and the USA (first period). The biggest differences within the first period were between the UK and China, and the UK and Other. The second scrape shows the biggest difference between Other and the UK, and China and the UK. The direction of differences in the first and the second scrape period remained the same. The UK seems to have the lowest similarity scores across scrape periods, while Other has the highest. Since product categories and origins significantly changed similarity scores, some products might be worth focusing on, such as Electronics and products from China.

|         | China  | Undec. | USA    | H. K.  | Other  | Europe | UK     |
|---------|--------|--------|--------|--------|--------|--------|--------|
| China   | -      | -0.82* | -0.11* | -1.29* | 0.17*  | -0.33* | -1.56* |
| Undec.  | 0.51*  | -      | 0.65*  | -0.49* | 1.08*  | 0.51*  | -0.80* |
| USA     | 0.14*  | -0.29* | -      | -1.14* | 0.25*  | -0.17* | -1.24* |
| H. K.   | 1.45*  | 0.88*  | 1.21*  | -      | 1.47*  | 0.96*  | -0.35* |
| Other   | -0.32* | -0.73* | -0.33* | -1.72* | -      | -0.56* | -1.58* |
| Europe  | 0.34*  | -0.17* | 0.12   | -1.10* | 0.71*  | -      | -1.11* |
| UK      | 1.81*  | 1.13*  | 1.28*  | 0.47*  | 1.76*  | 1.21*  | -      |

Table 6. Cohen's *d* effect sizes for product origins comparisons (row to column) for the first (lower-left diagonal half) and second (upper-right diagonal half) scrape period; Undec. = Undeclared; H. K. = Hong Kong; Significance level: * = p < 0.05 (Bonferroni corrected).

Lastly, we compare similarity differences between scrape periods, which showed a significant Cohen's *d* effect size of *d* = -0.16, indicating a slight decrease in similarity scores from the first to the second scrape period. Thus, the results suggest fewer good matches between cryptomarket and eBay products over time.

### 4.2. Finding the same products from cryptomarkets on eBay

We manually examined the top 50 cryptomarket and eBay product pairs and randomly selected pairings (excluding the top 50) for both scrape periods to examine how well the product matched.

#### 4.2.1. Product matches in the first scrape

The top 50 matched products of the first period contained 13 unique cryptomarket products, and the random sample contained 47 unique cryptomarket products (Table 7).

| Category    | Top 50 | Random 50 |
|-------------|--------|-----------|
| Clothes     | 0      | 8         |
| Watches     | 2      | 9         |
| Electronics | 3      | 1         |
| Jewellery   | 14     | 18        |
| Other       | 31     | 14        |

Table 7. Product categories across the top 50 and 50 randomly selected product matches (first scrape).

Electronic products contained Apple smartphones or (Apple) headphones. Jewellery contained mostly watches in the top 50, but for the random sample, these also included necklaces, earrings, rings, handbags, and bars of gold or silver. Products categorized as "Other" mostly contained shoes, specifically Nike shoes, but also one sweatshirt in the top 50. For the random sample, products also included wallets, earrings, wristbands, handbags, caps, and gold bars. Clothes were predominantly shirts but also contained jackets and a hoodie in the random sample. Based on manual inspections of the image and text for the top 50 matched products, we found eight matches that seemed identical (the other pairings are discussed below). Specifically, we found five Nike shoes on eBay, which all resembled the Nike shoes found on Darkode (Figure 4: CM-1;), two Apple smartphones resembling smartphones found on White House Market (Figure 4: CM-2), and one watch found on World Market (Figure 4: CM-3). Figure 4: eBay-1,2,3 shows image examples of the matching eBay products.

Product titles of cryptomarket and eBay pairs also exhibited high resemblance (Table 8), with only slight variations in word usage. The matching Nike shoe also shows the same brand or product identification number (Table 8: A). Prices are (substantially) lower for these cryptomarket products than for the matching eBay products.

Looking at the content of the descriptions for the cryptomarket products and the matching eBay products, we can see that a large portion of the text differed, covering various aspects of the product or warranty and shipment (See Appendix 7 for complete example descriptions). For example, cryptomarket descriptions often explain how to order, how long the shipment will take, what measures are in place to avoid detection, and how detections or complaints are handled. Besides shipping and warranty information, such aspects are mostly missing in eBay descriptions, as they are irrelevant concerns. However, both descriptions often contained additional product information, such as color options or sizes, which can be valuable in determining the similarity between the products. Given the images, titles, and descriptions, we can say that seven eBay products could be the same as those sold on the cryptomarket.

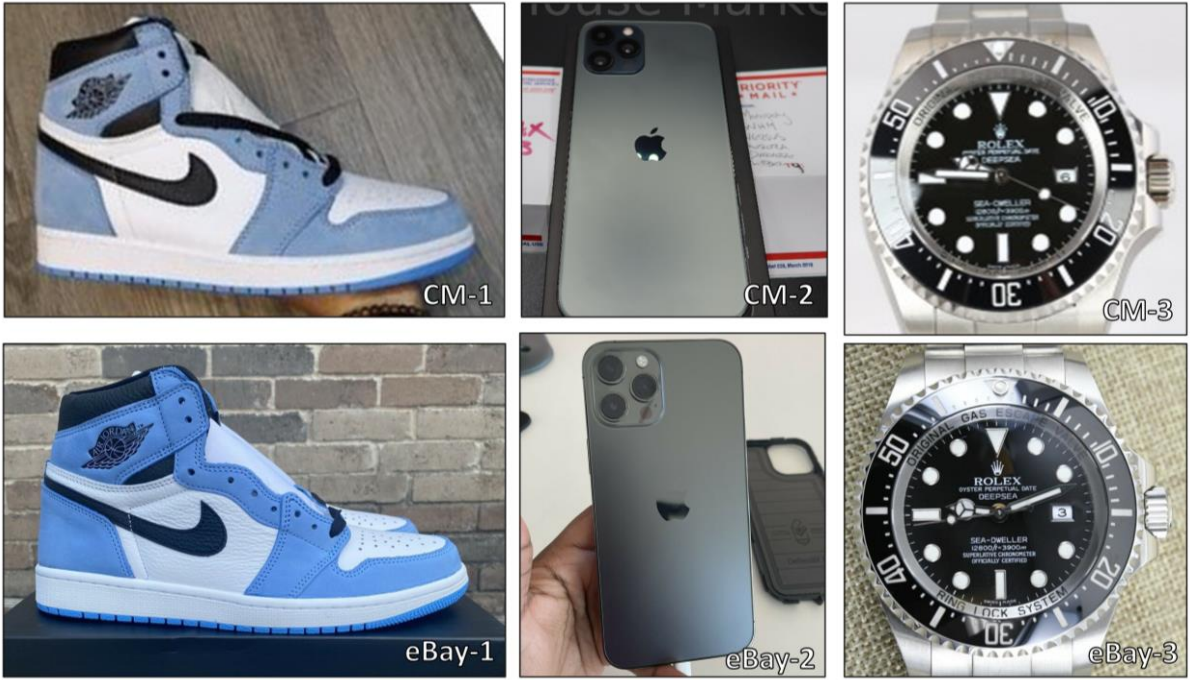

Figure 4. Examples of product images from the first scrape from CM and matching eBay products.

| | | Product type (figure reference) | | |
|---|---|---|---|---|
| | | Shoes (1) | Phones (2) | Watches (3) |
| Platform | CM Title | nike air jordan 1 retro high og 555088-134 | GRAPHITE - 512GB iPhone 12 Pro Max Sealed in Box - EZ BURN SERIES | Rolex - DEEPSEA SEA-DWELLER N V5S SAB 【UltimateAAA+】 |
| | CM Price | $238 | $300 | $300 |
| | eBay Title | Nike Air Jordan 1 Retro OG High White University Blue 555088-134 Men's Size 9.5 | iPhone 12 Pro Max - Verizon - 512GB - Graphite - Open Box | Rolex Deepsea Sea-Dweller 116660 44mm Watch |
| | eBay Price | $355 | $1,339 | $12,500 |

Table 8. Examples of titles and prices of matching products corresponding to the products in Figure 4; Words indicated with grey background were removed for the automated eBay web search.

Examining the random sample of 50 matches, we found two eBay products, with ranks 4,617 and 13,149, that seemed identical to cryptomarket products—specifically, a Louis Vuitton wallet and an iPhone. Although the images of the iPhone only contain the sealed box, making a visual comparison more difficult, the titles and descriptions match the phone model (version, memory, etc.).

### 4.2.2. Product matches in the second scrape

The second period contained 27 unique cryptomarket products, and the random sample of 50 contained 46 unique cryptomarket products (Table 9).

| Category | Top 50 | Random 50 |
|---|---|---|
| Clothes | 0 | 6 |
| Watches | 1 | 12 |
| Electronics | 5 | 1 |
| Jewellery | 40 | 17 |
| Other | 4 | 14 |

Table 9. Product categories across the top 50 and 50 randomly selected product matches (second scrape).

Electronic products were either Apple smartphones or (Apple) headphones. For the top 50, jewellery products included wristbands, rings, necklaces, earrings, and watches. The random sample also included rings, handbags, and bars of silver or gold. Products categorized as "Other" were mostly shoes, specifically Nike shoes and sweatshirts in the top 50, as well as pants, handbags, sunglasses, caps, and slippers in the random sample.

Based on the manual image and text inspections, three of the top 50 matched products seemed identical (Figure 5). Again, we found a match for the same Nike shoes (Figure 5: CM-1) previously identified in the first scrape, but also a match of a bag charm sold on White House Market (Figure 5: CM-2), and a match of a watch sold on World Market (Figure 5: CM-3). All three eBay matches are seen in Figure 5: eBay-1,2,3. The watch on the White House Market has a sticker on its glass, which is absent on the watch on eBay. However, such a sticker can most likely be removed for further sales.

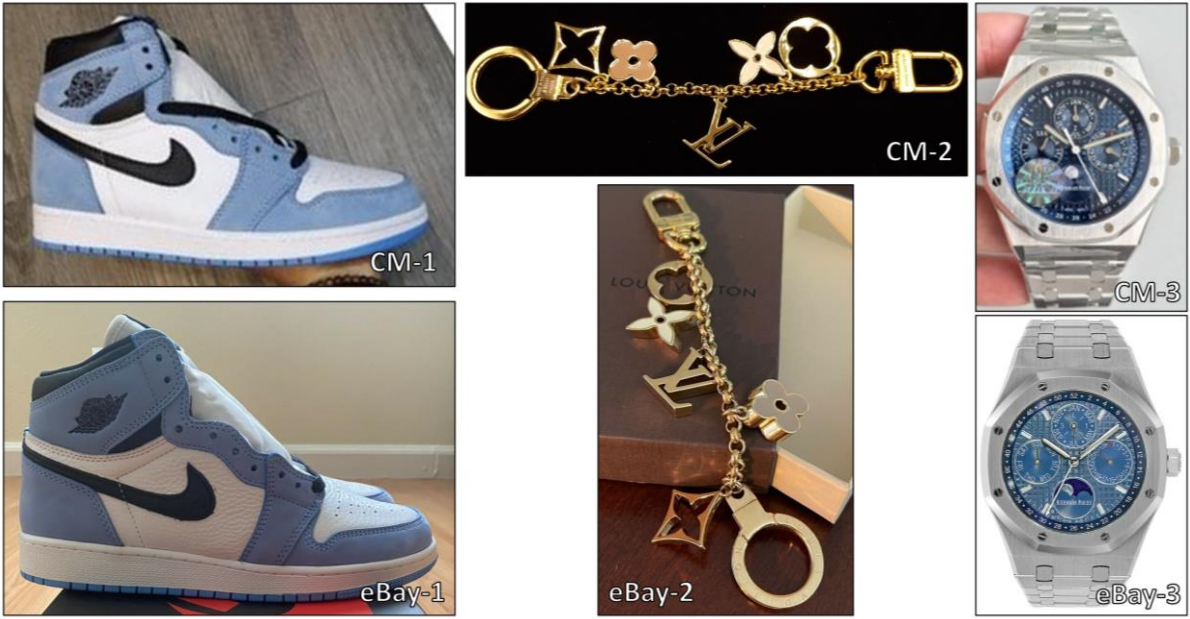

Figure 5. Examples of product images from the second scrape from CM and matching eBay products.

Again, product titles are very similar, and product prices are consistently lower on cryptomarkets (Table 10). However, the identification number of the Nike shoes within the titles does not match, which could be related to the indicated children's shoe size on the CM.

| | | Product type (figure reference) | | |
|---|---|---|---|---|
| | | Shoes (1) | Bag charms (2) | Watches (3) |
| Platform | CM Title | nike air jordan 1 retro high og 555088-134 | Louis Vuitton Bag Charm Chain Fleur de Monogram - RETAIL $830 - UNDETECTABLE | Audemars Piguet - ROYAL OAK PERPETUAL CALENDAR B 【UltimateAAA+】 |
| | CM Price | $238 | $300 | $450 |
| | eBay Title | Nike Air Jordan 1 Retro High OG GS University Blue 575441-134 Size 7Y | LOUIS VUITTON Bag Charm Chain Fleur de Monogram Key Ring (104623) | Audemars Piguet Royal Oak Perpetual Calendar Watch 26574OR.OO.1220OR.02 |
| | eBay Price | $350 | $750 | $215,997 |

Table 10. Examples of titles and prices of matching products corresponding to the products in Figure 6; Words indicated with grey background were removed for the automated eBay web search.

The corresponding product descriptions (Appendix 8) show a similar trend as previously identified but are more limited to factual details about the products (e.g., color, shape) and their shipment or warranty. Based on the images, titles, and descriptions, all three eBay products could be the same as advertised on the cryptomarket.

Examining the random sample of 50 matches, we found one eBay product, with the rank 1,245, that seemed identical to a cryptomarket product: a Rolex watch. The images, title, and descriptions match on both listings.

### 4.3. Highly ranked and similar product pairs that were not identical

32 products in the first and 12 in the second scrape period of the top 50 did not appear to be the exact same product but were highly similar and mainly varied only in their color scheme. For example, the corresponding cryptomarket Nike Shoes, seen in Figure 4: CM-1 and Figure 5: CM-1, were often matched with eBay Nike shoes, as seen in Figure 6. We also observed similar matching behavior for other products, such as the Audemars Piguet watch in Figure 5: CM-1 matching with the watch in Figure 6. The wristbands of the watches were often different, but these could easily be swapped for further resales. Thus, detecting the same product type with the same shapes or geometrical features seems to work better than accurate color detection. A possible reason for such color mismatches could be due to the color histogram comparisons. Since the color histograms account for the entire image, including the background of the products, the distribution of the histogram can be easily affected, possibly leading to inaccurate color comparisons of the products.

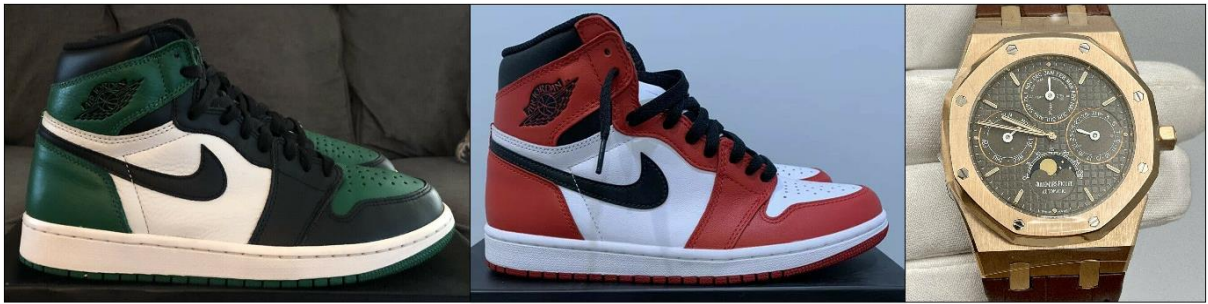

Figure 6. eBay products resembling counterfeits on cryptomarkets (Figure 4 and Figure 5: CM-1; Figure 5: CM-3) but with different color schemes.

### 5. Discussion

The current study uses information about cryptomarket counterfeits to find the same products on eBay. However, the system we tested cannot – and was not intended to – validate if the products found on eBay are, in fact, counterfeits. Instead, the purpose was to examine whether an automated approach could alleviate some of the workload current law enforcement agencies face and facilitate a better understanding of current trends in counterfeit sales. By partially automating an otherwise tedious web search, we can speed up the gathering of intelligence, which can be used further by manually inspecting highly similar products. Thus, the current system should not be regarded as a stand-alone solution to finding surface web counterfeits but rather as a partial automation of otherwise manual web searches.

## 5.1. How do product categories and origins affect product matching?

Based on the ANOVA analyses, we found that product similarities vary between product categories and product origins as well as the two combined. Specifically, post-hoc t-tests showed that similarities differed significantly between all product categories and almost all product origins. Finding the same products seemed more likely for Electronics and products originating from countries that contributed less than 1% of all origins ("Other"). In turn, finding the same products seemed least likely for clothes and products originating from the UK. A possible reason for good matches within electronics might be the detailed descriptions of specifications often not present in other product types (e.g., memory, model number, camera). We also observed that products originating from China had higher similarity scores than almost all other origins. Previous work shows that China is a predominant exporter of counterfeits, making the country a likely candidate to find very similar products (EUIPO, 2019; OECD/EUIPO, 2019), which is supported by our findings. However, it is unclear why finding highly similar or the same products differs for every product type and origin. Therefore, future research could investigate whether similarity cut-off scores, indicating which product matches should be manually examined, should be adjusted by product category or origins.

## 5.2. How does product matching change over time?

Our findings show that the overall similarity scores slightly decreased from the first to the second scrape period. Those results are also reflected in our manual inspections, in which we found more identical products in the first period than in the second. Although we found indications that eBay listings seemed to become less similar to cryptomarket listings over time, it is important to note that the most recent cryptomarket listings were collected (i.e., the top 50 in each category, sorted from newest to oldest) but without exact dates. Thus, we do not know precisely how long products were online, making a more accurate assessment of time effects difficult. For example, if we assume that counterfeits will appear first on cryptomarkets and then on the surface web, our current observations seem contradictory. However, our data collection could have captured a later stage of the product offer cycle, with products present for some time already and offers slowly decreasing over time (e.g., due to market saturation or sold-out products).

We also see scrape periods interacting with product categories and product origins. Specifically, we observed that compared similarities between most product categories diverge more from each other from the first to the second scrape, indicating greater differences between categories over time. However, comparing similarities between product origins, we observed a convergence across scrape periods for most comparisons, indicating closer similarities from the first to the second scrape period. Thus, we see opposing trends between product categories and product origins over time. Although the current exploratory observations of product changes over time are preliminary and need to be tested further, they give some indications that similarity scores behave differently over time depending on product groups and product origins. Thus, some specific products might show higher similarity for longer than other product types and might be worth tracking longer to find potential counterfeits.

### 5.3. Limitations

The current approach of collecting eBay data might strongly impact the matching procedure. Specifically, the default and opaque "best matches" settings in the eBay search query might be a bottleneck for finding good product matches. Many cryptomarket product names used for eBay searches are short and less descriptive, resulting in inaccurate search results. In contrast, the titles of good matches mostly contained some product details, such as exact model numbers. Automated systems for keyword generation, such as "KeyBERT" (Grootendorst, 2021), could be tested to generate more suitable search queries from lengthy product descriptions. Searches could be refined through price ranges or product specifics (e.g., size, color). However, prices can fluctuate, making specific price settings challenging to determine. Furthermore, observed cryptomarket listings might represent wholesales, complicating comparisons to single-item listings, or prices are only provided after customer inquiries, which is a recurring practice (Soska & Christin, 2015).

The current system finds some highly similar products across platforms, but we do not know how they relate to each other and if there are interdependencies. For example, products might be highly similar because the same vendor sells them; they might be resold by an individual who purchased the goods on a cryptomarket; or vendors on cryptomarkets may have researched surface web platforms to determine which products should be counterfeited and offered on cryptomarkets. Alternatively, product similarities might also originate from a complex relationship between manufacturers and sellers, or listing information (e.g., images, texts) may be copied from other advertisements without a direct connection between vendors. Thus, further research would have to be conducted to explore those possible interdependencies.

Another issue are the low agreement scores between annotators who rated the similarities between cryptomarket and eBay products. The human-rated similarity scores are essential in informing the regression model and determining how the individual automated scores should be weighted. Such agreements show the difficulty for humans to judge similarity but also jeopardize the meaningful training of a regression model, which was observed in the low model performance scores. However, the system found exact product matches and demonstrated utility. Nonetheless, more consensus between annotators is desirable, and future studies should examine reasons for possible solutions for the annotator's disagreements. For example, Annotators might need a revised version of instructions or similarity definitions. Furthermore, more annotators for each product pair might be required to find a more robust similarity rating. Alternatively, annotators may be invited to revise their ratings after others have rated the same item(s), applying an iterative process of refining the ratings, or raters may be asked to discuss items collectively rather than independently.

Furthermore, we observed some unexpected negative associations in the regression coefficients for text and image features, suggesting that some automated similarity scores negatively impacted the overall similarity score. However, despite these shortcomings, we observed a big difference qualitatively. That is, our comparison of the top 50 ranked pairings and a random sample of pairings suggest that the current system rankings can generally discriminate between better and worse product matches. Thus, despite the quantitative indications of poor performance (i.e., MAPE scores, negative regressions coefficients), the system showed utility qualitatively.

### 5.4. Future work

For this study, we collected product information from only some of the available counterfeits on cryptomarkets. Future studies could automate some collection processes to expand the collection of counterfeits. Instead of collecting all data from a market, which can be very time-consuming, the relevant product categories could be manually determined and provided to a scraper. Such an approach would expand the collection while avoiding the collection of irrelevant listings. Future studies could also collect data more frequently over a prolonged period (e.g., once a month over a year) to more accurately examine how offers change over time. The search for the same products could also be expanded to other platforms. Similarly, the vendors could be examined and compared across platforms. For example, whether vendors on cryptomarkets that sell various counterfeits show similarities to vendors on surface web platforms and if specific product types might be sold together.

While manually inspecting matched cryptomarkets and eBay products, we observed many pairs that showed strong resemblance in shapes and type but differences in their color scheme. Thus, accurate color detection between products seems difficult. The current matching through color histograms, which contain pixels of the entire image, including the background colors, might skew the distribution to an unfavorable comparison. Future approaches could test whether masking the background (i.e., separating the product from the background) could support a better color analysis. However, finding highly similar products with a different color scheme is also helpful. As for shoe sizes, which are often specified in the description, some cryptomarket products are also available in different colors, specified in the text but not visible in the example images. Although finding the same products is favorable, very similar products should not be discarded immediately to find potential counterfeits.

### 6. Conclusion

With the current work, we devised an automated system that finds eBay products similar to openly sold counterfeits on cryptomarkets. We found products across platforms that appear to be the same (although we do not know if the identified eBay products are counterfeits) or are very similar (e.g., differences in color schemes), which would warrant further inspection, as would the associated sellers. We also found some evidence suggesting that finding matching products across platforms becomes more difficult over time (at least for the two periods considered) and depends on product type and origins. Streamlining the integration of human judges for evaluating good product matches and finding suitable cryptomarkets would improve a future version of our current approach. Furthermore, finetuning the applied models to receive more robust similarity ratings and integrating the results within a graphical interface would make manual inspections more efficient. Thus, future versions of our approach could be used to investigate the possible connections between cryptomarket and surface web listings, as well as hold practical value in supporting the detection of counterfeits on the surface web.

**Appendix 1: Descriptive statistics of word occurrences**

| Listing texts | min | | max | | median | | mean | | SD | |
|---|---|---|---|---|---|---|---|---|---|---|
| | CM | eBay | CM | eBay | CM | eBay | CM | eBay | CM | eBay |
| **Title** | 2 | 1 | 25 | 22 | 5 | 12 | 5.6 | 11.19 | 2.92 | 2.62 |
| **Item specifics** | - | 0 | - | 690 | - | 97 | - | 108.23 | - | 52.97 |
| **Description 1** | 4 | 0 | 891 | 182 | 70 | 0 | 95.77 | 12.67 | 111.95 | 22.29 |
| **Description 2** | - | 0 | - | 65060 | - | 191 | - | 533.97 | - | 2762.94 |

Word distributions of cryptomarket (CM) and eBay listings texts, which are used in our analyses to generate text similarity scores between CM and eBay listings. Item specifics contain factual properties of the product (e.g., product height, weight, brand). eBay listings have two descriptions corresponding to a short (1) and long (2) text in most cases. Due to a change in the eBay webpage, most item specifics were not collected in the second scrape period. Thus, the statistical descriptors for item specifics only refer to the first eBay scrape period. 2,108 eBay listings were fully or partially not in English and were automatically translated with the Google translator in Python, using the package "deep translator" (Baccouri, 2020/2020).

**Appendix 2: Example of rescaled cryptomarket images**

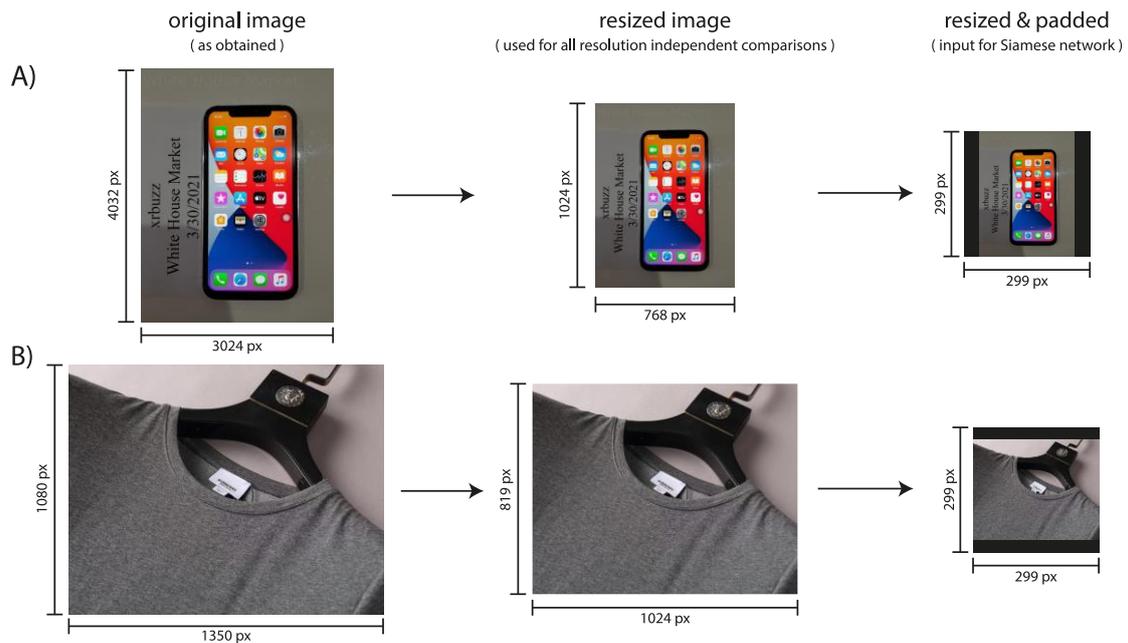

Example of rescaling procedure of cryptomarket (CM) images, restricting the largest dimension (width, height) to a maximum of 1024 px while preserving the original aspect ratio. Smaller images are not rescaled. All images (from both eBay and CM listings) are rescaled to 299 px x 299 px and padded, if necessary, to fit the input dimensions of the Inception V3 feature extractor backbone of the Siamese network. **A)** rescaled image from listing "SILVER iPhone 12 Pro Max 512GB - Sealed In Box". **B)** rescaled image from listing "Burberry mercerized cotton t-shirt 77005".

# Appendix 3: Siamese network architecture

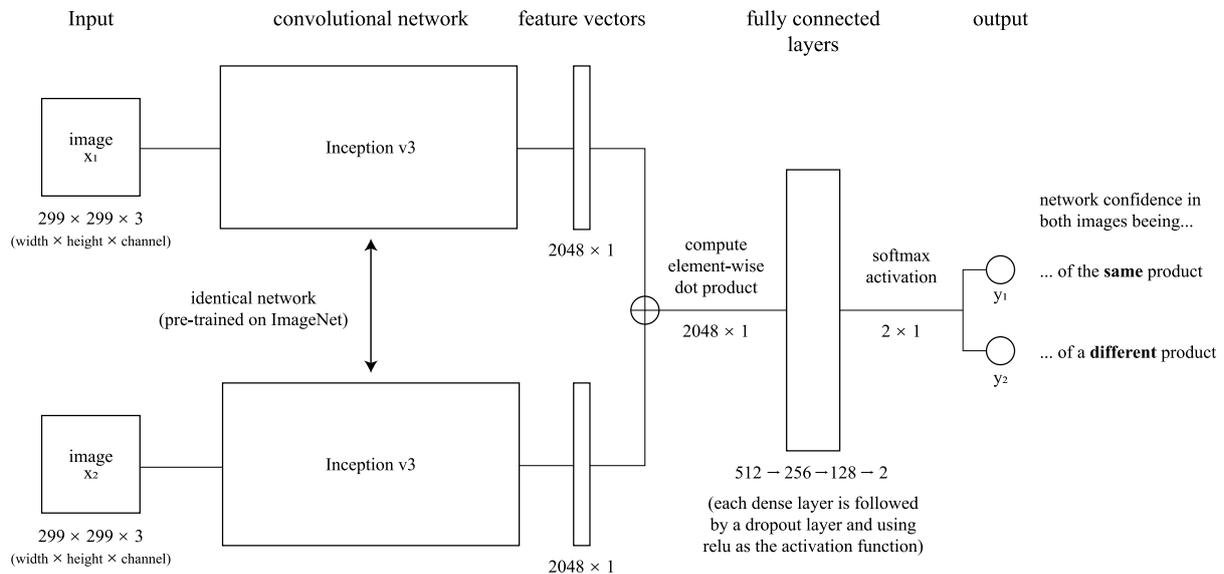

Depiction of the flow of information within the Siamese network architecture. The values below each element indicate its dimensions. The inputs are made up of two images to be compared: $x_1$, an image from a cryptomarket listing to $x_2$, an image from a surface web listing. We use an Inception V3 network, pre-trained on the ImageNet dataset, to produce high-level feature vectors from both input images. The element-wise dot product of the two produced feature vectors is computed, and the resulting vector is passed to a set of fully connected layers, which produce the final output. The network forms two predictions, $y_1$ and $y_2$, the similarity and dissimilarity, respectively, of the two images and can therefore be trained as a binary classifier.

# Appendix 4: Detailed similarity score calculation

The fully connected network head of the Siamese architecture was trained using the ADAM optimizer (Kingma and Ba 2014) over a total of one million iterations with a batch size of 16. To compute the final similarity scores, instead of noting the binarized output of the network, we use the SoftMax output of both output nodes as normalized measures for similarity. Thus, we received two output values in the form of the activation of the output nodes, indicating the network's prediction and confidence regarding whether the input images were of the same identity. On the one hand, a high activation (values close to 1) of the first output node and a low activation (values close to 0) of the second node indicate high confidence in the images being of the same identity. On the other hand, a low activation of the first node and a high activation of the second node indicates the opposite. If the activation of either or both nodes is close to 0.5, the confidence in the prediction is low. To simplify the integration of the two scores with the other metrics, we combined them by subtracting the dissimilarity from the similarity score leading to a single score ranging from -1 (low similarity, high confidence), over 0 (low confidence), to 1 (high similarity, high confidence).

**Appendix 5: Detailed description of sampling annotation data**

To determine a single similarity score for each product pair, we first created a single score describing the similarity of the title, description, and images by merging the different similarity metrics. We aimed to ensure equal weightings for the three aspects by obtaining a single score for each listing attribute (i.e., title, description, images). For each listing attribute, we determined if the score for a metric (e.g., colour histogram) given to a comparison pair was unusually high (or low, depending on the meaning of the score) by calculating if the score value was above (or below) two standard deviations from the mean of all scores for that metric. To avoid possible product-specific scoring biases, we calculated the means and standard deviations for each product group (watches, shoes, etc.) separately and determined whether each product's similarity value deviated unusually from its product group. The merged similarity score for any given product pair attribute consists of the count of unusual deviations (2 · STD ± Mean) of each metric. We then normalized each merged score, resulting in three scores for the title, description, and images, each ranging from 0 (low similarity) to 1 (high similarity). By standardizing the scores, we counteract the unequal number of metrics between text and image scores. eBay listings contain several subsections of detailed information (e.g., product specifics), which were all merged into one text labelled as product description.

We scored all cryptomarket and eBay (first scrape only) product pairs and ranked them from the highest to lowest similarity scores. To determine if the ranking could capture a preliminary order of similarity, we inspected the top and last 50 ranks of product pairs. Within the top 50 ranks, we found that products were correctly matched on product types, including four products that seemed to depict and describe the same product (2 shoes, a shirt, and a bag chain). Within the last 50 ranks, products seem to match very poorly, including mismatches of product types (e.g., a shoe with a shirt). Since the ranked products appeared to follow a somewhat sensible order, we sampled the top and last 250 products and a random sample of 500 from the remaining products, resulting in a sample of 1000 product pairs.

**Appendix 6: Regression models settings**

| Model | Settings |
|---|---|
| OLS | *fit_intercept=True, normalize='deprecated', copy_X=True, n_jobs=None, positive=False* |
| SVR | *kernel='linear', degree=3, gamma='scale', coef0=0.0, tol=0.001, C=1.0, epsilon=0.1, shrinking=True, cache_size=200, verbose=False, max_iter=- 1* |
| RFR | *n_estimators=100, criterion='squared_error', max_depth=None, min_samples_split=2, min_samples_leaf=1, min_weight_fraction_leaf=0.0, max_features='auto', max_leaf_nodes=None, min_impurity_decrease=0.0, bootstrap=True, oob_score=False, n_jobs=None, random_state=319, verbose=0, warm_start=False, ccp_alpha=0.0, max_samples=None* |

**Appendix 7: Full product descriptions (eBay scrape period 1)**

| Fig | Description |
|---|---|
| 4: CM-1 | Nike Air Jordan 1 Retro High OG 555088-134<br>Size: 36 36.5 37 38 38.5 39 40 40.5 41 42 42.5 43 44 44.5 45 46<br>The popularity of Air Jordan 1 this year has remained constant, but Jordan Brand has never stopped developing its new color scheme. The Jordan 1 High OG (University Blue) will be one of the first versions of the Jordan brand in early 2021. This Air Jordan 1 is matched with white, college blue and black colors. Although the images have not yet been leaked, they are expected to have white leather on their uppers, while college blue is on the overlay. Other details will include black trim, white midsole and blue rubber outsole. |
| 4: eBay-1 | Item specifics Condition: New with box: A brand-new, unused, and unworn item (including handmade items) in the original packaging (such as the original box or bag) and/or with the original tags attached. See all condition definitions - opens in a new window or tab ... Read more about the condition Pattern: Colorblock Model: Air Jordan 1 Style Code: 555088-134 Style: Sneaker Character: Michael Theme: University US Shoe Size: 9.5 Features: Comfort, Cushioned Color: Blue Country/Region of Manufacture: China Silhouette: Jordan 1 Retro High OG University Blue Upper Material: Leather Brand: Jordan Idset_Mpn: 555088-134 Department: Men Performance/Activity: Basketball Type: Athletic Product Line: Jordan<br><br>Nike Air Jordan 1 Retro OG High White University Blue 555088-134 Men's Size 9.5. Condition is "New with box". Shipped with FedEx Ground or FedEx Home Delivery. |
| 4:<br>CM-2 | APPLE SETUP WHEN YOU FIRST RUN IT:<br>APP STORE IS FUNCTIONAL:<br>ITUNES IS FUNCTIONAL:<br>FLASHED TO MAKE ALL DETAIL LOOK LIKE APPLE:<br>SECRET MENU:<br>ABILITY TO CHAMGE YOUR IMEI:<br>ABILITY TO CHANGE YOUR MEID:<br>ABILITY TO CHABGE YOUR SERIAL NUMBER:<br>PRE-ROOTED:<br>ABILITY TO CLONE AND RECIEVE MESSAGES AS ANYONE IN THE WORLD:<br>ABILITY TO BURN PHONE ON A MOMENTS NOTICE:<br>ABILITY TO MAKE YOU RICH:<br>GREAT FOR CARDING:<br>GREAT FOR ANONYMITY:<br>IN YOUR CART CAUSE YOU ARE BRILLIANT:<br>GREAT CUSTOMER SERVICE WITH AN ACE OF SPADES UP HER SLEEVE JUST IN CASE<br>YOU CAN'T PULL OFF A SALE IN 30 DAYS:<br>DEDICATED SUPPORT TEAM:<br><br>Put this in your cart, let's make some money<br><br>Just so you know from jump I have a video of an unboxing that we took in house. Taken by a Professional!!!......pothead but this dude ain't bad looks....like it was done by someone in at least...3rd grade? I have<br>absolutely no problem taking one for you as well if needed. Ask me to perform certain tasks on it so you can see if it is right for you.<br><br>Unboxing:<br>You would never in your life know from opening this thing that it is a clone. Ive opened them side by side. No difference.<br><br>Turn it on:<br>Same startup screen , same layout, same apps. These are the upgraded versions too. So no lag whatsoever, its smooth as ever.<br><br>Take a picture:<br>Fantastic quality<br><br>Security:<br>Face Id<br>Fingerprint unlock<br><br>Insert SIM:<br>IT WORKS! same imei numbers as actual iPhone. Most are valid. You will not get a dud from us with these 12's. They are fresh and we will send a picture of the imei for each buyer to test before shipment to see with their own eyes<br><br>It's all the same....but the rice....waaaaaay less<br><br>This shit IS an iPhone..just one that was dropped on its head as a baby and is hooked on droids.... Thats right it is run with android. but ill tell you this - quick handoffs ---hahaha that's what we used to say, FUCK THAT, grab a beer with the dude, go to grandmas for your special<br>holiday and let em pass it around to the whole family to test before purchase..<br><br>You will receive a nice little package and ill give you a little instruction booklet on how to be really good at pushing these out. honestly the last thing you will ever see of these people will be smiling faces. I'm not exaggerating in the slightest when I say the last guy I hooked up SKIPPED away from me he was so happy. Hah. shit. I just dont understand what people see in these apple products. It looks like a bigger more fucked up version of the 5. IT NEVER CHANGES. But whatever these things are FLYING off the handle at $12-$1800 even broken they are worth 8 or 9.<br><br>Shit BUY 10 break ALL of em and give it for six and youll be rolling in it. Thats THREE THOUSAND DOLLARS JUST FOR BROKEN IPHONES. you know people woll be looking for parts, a screen, a camera WHATEVER, that could be your hustle and nobody would be the wiser. |

| | | |
|---|---|---|
| | | Personally I dont have the time nor the poker face for it so this is what I do. BUT my lovely staff here have each had to get rid of our base models. Part of their interview was to sell 5 of the ones worse than these so if you need ANY help trust me they know what the fuck, where<br>the fuck, how the fuck to pull it off.<br><br>I'll be putting up Note20s and Note9s soon. Those are beautiful little creatures as well.<br><br>I refuse to sell the Note10's cause someone fucked up down the line didnt make the screen edge to edge and decided it was acceptable. We only want to give you guys the best.<br><br>If you need more than one, the price breaks down nicely. Just shoot us a message and we will throw up a personalized listing.<br><br>Oh and the package includes:<br><br>1x 512GB iPhone 12 Pro Max<br>1x iPhone case<br>1x iPhone screen Protector<br>1x Lightening to USB-C Authentic Apple Charging Cord<br>1x Apple welcome packet<br>1x Sim Card Pin<br>1x iPhone OEM Box<br><br>Enjoy |
| 4: eBay-2 | | Item specifics Condition: Open box : An item in excellent, new condition with no wear. The item may be missing the original packaging or protective wrapping, or may be in the original packaging but not sealed. The item includes original accessories. The item may be a factory second. See the seller's listing for full details and description. See all condition definitions - opens in a new window or tab Seller Notes: "Open Box Condition: Brand new, but box has been opened." Brand: Apple Connectivity: 5G, Lightning, Bluetooth, NFC, Wi-Fi Model: Apple iPhone 12 Pro Max Processor: Hexa Core Style: Bar Operating System: iOS Storage Capacity: 512 GB Manufacturer Color: Graphite Features: Proximity Sensor, Barometer, LiDAR Scanner, Accelerometer, Fingerprint Sensor, E-compass, Ambient Light Sensor, Gyro Sensor, eSIM Camera Resolution: 12.0 MP Color: Gray MPN: MG9K3LL/A Network: Verizon Screen Size: 6.7 in UPC: 0194252019887 EAN: 0194252019887<br><br>Search our eBay store 🔎 iPhone 12 Pro Max - Verizon - 512GB - Graphite - Open Box Device Details ⚙ Network: Verizon - Can only be used with Verizon. Is not unlocked for use with other networks. Brand: Apple Model: iPhone 12 Pro Max Storage: 512GB Color: Graphite What's In The Box 📦 iPhone 12 Pro Max Box & All Manufacturers Sealed Accessories Condition 🏅 🏅 🏅 Certified - Open Box: Brand new, but box has been opened. What is a certified device? Before we clear a device for resale it must pass a series of functional, database and cosmetic inspections. These inspections are performed by our expertly trained team of in-house technicians using state of the art software to ensure there are no testing errors. 60 Day - Free Returns If you aren't completly satisfied with your purchase you can return it for a refund or exchange. We'll even cover the return shipping. 1 Year warranty 🎗 This device comes with a 12 Month (365 day) warranty starting at the date of purchase. If your device stops working properly from normal use during this timeframe you can ship it back to us and we'll either repair your device, send an exchange, or send a refund. Additional Details This warranty does not cover accidental damage of any kind. This warranty is void if any repairs or modifications have been performed or attempted to device hardware or software. The remedy for your warranty claim will be decided at our sole discretion. Generally speaking, we'll attempt to repair your device or send an exchange before considering a refund. This warranty only extends to the original purchaser; it cannot be transferred if the device is sold / given to a 3rd party. Why Buy From Us? Based in Sunny Florida, [Anonymized name] has been helping people all over the USA buy and sell gadgets since 2016. If you do a quick search for our company you'll see that we have great reviews not only on eBay, but all over the web. Our customers love us for one simple reason - we deliver on promises. Our Promises Quality Devices - We don't sell junk. Our certified devices have been tested extensively to make sure they'll work perfectly right out of the box. Fast Shipping - All orders are shipped the same or next business day after purchase. Orders placed before 2PM EST Tuesday - Friday will be shipped the same day. Amazing Customer Service - Our philosophy on customer service isn't a new one, we treat people how we like to be treated. That means fast & accurate communication, no-hassle 60 day free returns and a 1 year warranty. |

| | |
|---|---|
| 4: CM-3 | Brand---Rolex quality level---UltimateAAA+ Manufacturer:N Window material:Sapphire glass Bezel material:ceramics Case material:stainless steel Strap material:stainless steel Case Diameter:44 MM waterproof:60 M Movement:ETA 2836 Automatic mechanical movement. Vibration frequency:28800 function:Date.hour.minute.second. Dial luminous:Yes ---more images--- [Anonymized link] ---box images--- [Anonymized link]  If you want to buy a box. the certificate please choose 100USD in the shipping option. Because the box is large.the shipping is expensive. When you buy a watch. please fill in the address information in this format: name: address: City/State: Postcode: country: 1. When you order a watch, I expect to send the order within 4-8 days. I need time to order the watch and check the quality. When I send an order, I will provide tracking number information. 2. The United States, Canada, and Australia usually deliver in 10-15 days. EU, UK usually deliver in 10-20 days. 3. I guarantee that the goods can be delivered successfully. If the goods are lost or detained by the customs, I will bear the loss. |
| 4: eBay-3 | Item specifics Condition: Pre-owned: An item that has been used previously. The item may have some signs of cosmetic wear, but is fully operational and functions as intended. This item may be a floor model or store return that has been used. See the seller's listing for full details and description of any imperfections. See all condition definitions - opens in a new window or tab ... Read more about the condition Water Resistance: 3900 m (390 ATM) Water Resistance Rating: Diver \'s 300 m (ISO 6425) Model: Sea-Dweller Band Material: Stainless Steel Country/Region of Manufacture: Switzerland Type: Wristwatch Watch Shape: Round Features: Date Department: Men Customized: No Style: Casual, Sport, Diver, Luxury With Papers: No Case Color: Silver Year Manufactured: 2009 Caseback: Solid Indices: Non-Numeric Hour Marks, Round Indexes MPN: 116660 With Original Box/Packaging: No Hour Markers: Dot, index Dial Color: Black Case Material: Stainless Steel Band Color: Silver Gender: Men Reference Number: 116660 Buckle Type: Folding buckle Display: Analog Box/Papers: Box and Papers Brand: Rolex Movement: Mechanical(Automatic) Case Size: 44 mm Warranty: 2-Year Watchbox UPC: Does not apply<br><br>[Anonymous name] Jewelry and Loan [Anonymous name] [anonymous rating] Sign up for newsletter Search within store Visit Store: [Anonymous name] Jewelry and Loan Items On Sale Categories Other<br><br>Rolex Deepsea Reference 116660 Stainless steel 44mm Beast. This is a great watch in excellent condition with no box or card. Ready for your wrist with speedy delivery. We Guarantee Authenticity on all Merchandise. We Are A Long Standing Member of the [Anonymous name] [Anonymous name] in operation since [Anonymous date]. Two locations [Anonymous location] and [Anonymous location] WE ONLY ACCEPT PAY-PAL. AND CREDIT CARDS THROUGH PAYPAL. PLEASE SERIOUS inquiries ONLY FOR ANY FURTHER ASSISTANCE WITH PURCHASED ITEMS PLEASE CONTACT Attn. Zef WITH PROPER RETURN INSTRUCTIONS AND ADDRESS (Buyer is responsible for return shipping) We try to ship within 1 to 3 business days from the payment. Buyer pays a fixed rate for shipping and handling for the United States and International sales Payments must be made in 2 days of purchase. REMINDER : WE DO NOT SHIP TO "FPO/APO, OR" PO BOX "LOCAL PICK UP IS ACCEPTED AND REQUIRES THE CUSTOMER TO PAY 6% SALES TAX OF TOTAL AMOUNT AT THE TIME OF PICK UP" ONLINE ITEMS ARE NOT ELIGIBLE FOR [Anonymous name] UPGRADE AND TRADE IN POLICY Thank You |

## Appendix 8: Full product descriptions (eBay scrape period 2)

| Figure | Description |
|---|---|
| 5: CM-1 | Nike Air Jordan 1 Retro High OG 555088-134<br>Size: 36 36.5 37 38 38.5 39 40 40.5 41 42 42.5 43 44 44.5 45 46<br>The popularity of Air Jordan 1 this year has remained constant, but Jordan Brand has never stopped developing its new color scheme. The Jordan 1 High OG (University Blue) will be one of the first versions of the Jordan brand in early 2021. This Air Jordan 1 is matched with white, college blue and black colors. Although the images have not yet been leaked, they are expected to have white leather on their uppers, while college blue is on the overlay. Other details will include black trim, white midsole and blue rubber outsole. |
| 5: eBay-1 | Nike Air Jordan 1 Retro High OG GS University Blue 575441-134 Size 7Y. |
| 5: CM-2 | Louis Vuitton Bag charm<br>Chain Fleur de Monogram<br>M65111<br><br>CONDITION IS NEW<br><br>COMES WITH:<br><br>1x LV CHARM |

| | | |
|---|---|---|
| | | 1x LV DUST BAG<br>1x CERTIFICATE OF AUTHENTICITY<br>1x CARE BOOKLET<br>1x LV BOX<br>1x LV SHOPPING BAG |
| 5: eBay-2 | | Louis Vuitton Bag Charm/Key Chain Fleur de Monogram Gold Plated. This is preowned in excellent condition with no signs of use or wear. Comes with box. Buy in US to avoid duty and taxes that I have already paid!! Please message for more info or photos. TY! |
| 5: CM-3 | | Brand---Audemars Piguet quality level---UltimateAAA+ Manufacturer:JF Window material:Sapphire glass Bezel material:stainless steel Case material:stainless steel Strap material:stainless steel Case Diameter:41MM waterproof:30 M Movement:Clone Automatic mechanical movement. Vibration frequency:28800 function:calendar.Date.hour.minute.second. Dial luminous:Yes ---more images--- [anonymous link] images--- [anonymous link] If you want to buy a box. the certificate please choose 160USD in the shipping option. Because the box is large.the shipping is expensive. When you buy a watch. please fill in the address information in this format: name: address: City/State: Postcode: country: 1. When you order a watch, I expect to send the order within 4-8 days. I need time to order the watch and check the quality. When I send an order, I will provide tracking number information. 2. The United States, Canada, and Australia usually deliver in 10-15 days. EU, UK usually deliver in 10-20 days. 3. I guarantee that the goods can be delivered successfully. If the goods are lost or detained by the customs, I will bear the loss. |
| 5: eBay-3 | | Audemars Piguet Audemars Piguet Royal Oak Perpetual Calendar Watch 26574OR.OO.1220OR.02 Details Department Unisex Adult Dial Color Blue Dial Pattern Grande Tapisserie Case Size 41 mm Customized No Case Material Pink Gold Seller Warranty Yes Warranty 5 Year Warranty Reference Number [anonymous number] Water Resistance 20m (2 ATM) With Papers Yes Features Perpetual Calendar, Sapphire Crystal Case Color Pink Gold Item description 41 mm 18K pink gold case, 9.5 mm thick, glareproofed sapphire crystal back, screw-locked crown, glareproofed sapphire crystal, blue dial with Grande Tapisserie" pattern, pink gold applied hourmarkers and Royal Oak hands with luminescent coating, blue inner bezel, Manufacture 5134 selfwinding movement with perpetual calendar with week indication, day, date, astronomical moon, month, leap year, hours and minutes, approximately 40 hours of power reserve. Water resistant to 20 meters. Condition: Used• FREE Overnight Express & Insured FedEx Domestic & International overnight shipping. • International import duties, taxes, and charges are not included in the item price or shipping cost. These charges are the buyer's responsibility. • Full 5-YEAR warranty If the item fails because of a manufacturer's defect. [anonymous name] will repair or replace the item at absolutely no cost to you. Warranty coverage except: Abuse of the timepiece, accidental water intrusion caused by user, outside modifications and third-party repair attempts of any kind will void the warranty. External damage to the product not covered under warranty: damage resulting from abusive wear, crystal/glass, watch bracelet, watch bezel, straps, screws. • Please feel free to contact us directly if you have any questions. We are happy to assist with any inquiry. |